# Crystal-symmetry-paired spin-valley locking in a layered room-temperature antiferromagnet

Fayuan Zhang[1*], Xingkai Cheng[2*], Zhouyi Yin[3*], Changchao Liu[4*], Liwei Deng[5], Yuxi Qiao[5], Zheng Shi[2], Shuxuan Zhang[3], Junhao Lin[3], Zhengtai Liu[6], Mao Ye[6], Yaobo Huang[6], Xiangyu Meng[6], Cheng Zhang[7], Taichi Okuda[8], Kenya Shimada[8], Shengtao Cui[9], Yue Zhao[3#], Guang-Han Cao[4#], Shan Qiao[5#], Junwei Liu[2#] and Chaoyu Chen[1,10#]

[1] Shenzhen Institute for Quantum Science and Engineering, Southern University of Science and Technology, Shenzhen, China.

[2] Department of Physics, The Hong Kong University of Science and Technology, Hong Kong, China.

[3] Department of Physics and Shenzhen Institute for Quantum Science and Engineering, Southern University of Science and Technology, Shenzhen, China.

[4] Department of Physics, Zhejiang University, Hangzhou, China.

[5] Shanghai Institute of Microsystem and Information Technology, Chinese Academy of Sciences, Shanghai, China.

[6] Shanghai Synchrotron Radiation Facility, Shanghai Advanced Research Institute, Chinese Academy of Sciences, Shanghai, China

[7] Graduate School of Advanced Science and Engineering, Hiroshima University, Higashi-Hiroshima, Hiroshima, Japan.

[8] Research Institute for Synchrotron Radiation Science (HiSOR), International Institute for Sustainability with Knotted Chiral Meta Matter (WPI-SKCM2) and Research Institute for Semiconductor Engineering (RISE), Hiroshima University, Higashi-Hiroshima, Hiroshima, Japan.

[9] National Synchrotron Radiation Laboratory, University of Science and Technology of China, Hefei, China

[10] Institute of Advanced Science Facilities, Shenzhen, China

[*] These authors contributed equally to this work.

[#]Correspondence should be addressed to Y.Z. (zhaoy@sustech.edu.cn), G.C. (ghcao@zju.edu.cn) S.Q. (qiaoshan@mail.sim.ac.cn), J.L. (liuj@ust.hk) and C.C. (chency@sustech.edu.cn).

**Abstract:**

Recent theoretical efforts predicted a type of unconventional antiferromagnet characterized by the crystal symmetry *C* (rotation or mirror), which connects antiferromagnetic sublattices in real space and simultaneously couples spin and momentum in reciprocal space. This results in a unique *C*-paired spin-valley locking (SVL) and corresponding novel properties such as piezomagnetism and noncollinear spin current even without spin-orbit coupling. However, the unconventional antiferromagnets reported thus far are not layered materials, limiting their potential in spintronic applications. Additionally, they do not meet the necessary symmetry requirements for nonrelativistic spin current. Here, we report the realization of *C*-paired SVL in a layered room-temperature antiferromagnetic compound, $Rb_{1-\delta}V_2Te_2O$. Spin resolved photoemission measurements directly demonstrate the opposite spin splitting between *C*-paired valleys. Quasi-particle interference patterns reveal the suppression of inter-valley scattering due to the spin selection rules, as a direct consequence of *C*-paired SVL. All these experiments are well consistent with the results obtained from first-principles calculations. Our observations represent the first realization of layered antiferromagnets with *C*-paired SVL, enabling both the advantages of layered materials and possible control through crystal symmetry manipulation. These results hold significant promise and broad implications for advancements in magnetism, electronics, and information technology.

## I. Introduction

The realization and control of spin-polarized electronic states in solids represent a crucial step towards spintronics for encoding and processing information[1-3]. Typically, spin polarization is generated by coupling an electron's spin to other degrees of freedoms such as orbital[4-6] or magnetic moment[7]. The former employs the spin–orbit coupling (SOC) and generates momentum-dependent spin splitting in crystals lacking inversion-symmetry, known as the Rashba–Dresselhaus effect[8,9]. The latter is traditionally associated with the time-reversal symmetry breaking from the internal magnetization of ferromagnets, resulting in momentum-independent Zeeman-type spin splitting in the band structure.

Recent theoretical efforts have predicted unconventional antiferromagnetic (AFM) crystals that exhibit momentum-dependent spin splitting without SOC or net magnetization[10-14]. These antiferromagnets, termed altermagnets[13,15], are unique because their crystal symmetry $C$ (e.g., rotation or mirror) connects opposite-spin magnetic sublattices in real space while simultaneously couples spin and momentum in reciprocal space, leading to the new $C$-paired spin-valley locking (SVL)[10]. For conventional SVL in transition metal dichalcogenides, valleys with contrasting spin splitting are connected by time reversal symmetry $T$: $E_\uparrow(K) = E_\downarrow(-K)$, while strong SOC and inversion breaking are required to split bands[16]. This is referred to as $T$-paired SVL (Fig. 1a1). In contrast, for $C$-paired SVL systems, spin-polarized valleys are connected by crystal symmetry $C$: $E_\uparrow(K) = E_\downarrow(CK)$ (Fig. 1a2), where spin splitting arises from the exchange coupling between electrons and AFM orders, which can be as large as the scale of eVs[15,17].

Compared to $T$-paired SVL, $C$-paired SVL enables novel physical responses, such as unconventional piezomagnetism[10] and large noncollinear spin current[10,18-20] including both nonrelativistic spin-polarized currents as in ferromagnetic materials due to unequal spin-up and spin-down carriers and nonrelativistic pure spin currents like in strong SOC materials due to the spin Hall effect. These properties facilitate both static and dynamical controls of spin and valley degrees of freedom. However, nonrelativistic spin currents rely on conductance anisotropy for both spin channels[10,20], imposing restrictions on the choice of crystal symmetries. Specifically, anisotropy vanishes if $C_n$ ($n \geq 3$) exists in the same sublattice (see Supplementary Section I for details). Despite extensive theoretical and experimental efforts to explore unconventional antiferromagnet based on emerging materials such as $\alpha$-MnTe[21-23], CrSb[11,24], MnTe$_2$[13,17,25] and RuO$_2$[26-34], none of them satisfies the symmetry and conductivity requirements to realize nonrelativistic spin-conserved spin current due to the altermagnetism. As summarized in Table I, among these materials, each magnetic sublattice of $\alpha$-MnTe and CrSb possesses $C_3$ symmetry that leads to isotropic conductance for both spin channels and results in non-polarized current. MnTe$_2$ is semiconducting and hence it can realize both spin-polarized current and pure spin current, while the spin is not conserved due to its noncoplanar magnetic structure. In addition, the low critical temperature (87 K) of MnTe$_2$ further hinders its practical applications. As for RuO$_2$, it remains controversial whether its ground state is antiferromagnetic[30,31] or nonmagnetic[29], despite evidence of anomalous Hall effect[26,27] and spin splitting[28,35].

In addition, all these unconventional antiferromagnets reported are not layered materials, which limits their potential for exfoliation from bulk or integration with other materials to control their properties at the microscopic level. This restriction hinders the exploration of various effects in 2D materials that have been discovered in recent years, for example, the realization of topological superconductors via superconducting proximity effect[36], tunable electronic properties through gating[37,38], modification of band structures through strain[39] and potential to form twisted systems like moiré superlattices[40]. Therefore, exploring layered materials in altermagnets is highly anticipated to develop high-density, high-speed and low-energy-consumption spintronic devices.

In this work, employing spin and angle-resolved photoemission spectroscopy (Spin-ARPES), scanning tunnelling microscopy/spectroscopy (STM/STS) and first-principles calculations, we demonstrate unambiguously the realization of $C$-paired SVL in a layered, room-temperature AFM compound, Rb$_{1-\delta}$V$_2$Te$_2$O. Image-type Spin-ARPES[41] measurements directly reveal the out-of-plane ($S_z$) spin polarization for all the three bulk Fermi pockets at the tetragonal Brillouin zone (BZ) boundary. The sign of spin polarization for these Fermi pockets are opposite between adjacent X and Y valleys connected by crystal symmetry $C$, characteristic of $C$-paired SVL. Temperature dependent ARPES measurements demonstrate the persistence of SVL up to room temperature, in line with the AFM phase transition temperature. Furthermore, the quasi-particle interference (QPI) pattern from STM map reveals the suppression of inter-valley scattering due to the spin selection rules, serving as the direct physical consequence of SVL and weak SOC. All these results are well consistent with the results obtained from the first-principles calculations. Our work realizes the first layered room temperature AFM metal with alternating magnetic sublattice and a new type of spin splitting effect, providing an ideal material platform for further fundamental study and in device applications of spintronics and valleytronics.

Table I. Unconventional AFM materials with experimentally reported and potential properties.

| | Crystal Symmetry | SPC | PSC | Conductor | Collinear | Layered | $T_N$ (K) | Ref. |
|---|---|---|---|---|---|---|---|---|
| **$Rb_{1-\delta}V_2Te_2O$** | $\{C_{2x}\|\|C_{4z}\}$ $\{C_{2x}\|\|M_{xy(\bar{x}y)}\}$ | **Y** | **Y** | **Y** | **Y** | **Y** | **≥307** | **This work** |
| $RuO_2$ | $\{C_{2x}\|\|C_{4z}\}$ $\{C_{2x}\|\|M_{x(y)}\}$ | Y | Y | Y | ? | N | ? | 26-31 |
| α-MnTe | $\{C_{2z}\|\|C_{6z}\}$ $\{C_{2z}\|\|M_{y(z)}\}$ | N | N | S | Y | N | 310 | 21-23 |
| CrSb | $\{C_{2x}\|\|C_{6z}\}$ $\{C_{2x}\|\|M_{y(z)}\}$ | N | N | Y | Y | N | 704 | 11,24 |
| $MnTe_2$ | $\{C_{3(111)}\|\|C_{3(111)}\}$ $\{C_{2y(z)}\|\|M_{y(z)}\}$ | Y | Y | S | N | N | 87 | 13,17,25 |

Abbreviation: Crystal symmetry: spin and spatial rotation part of spin group symmetry connecting different AFM sublattices. SPC: spin polarized current (without SOC and net magnetization), PSC: pure spin current (without SOC and net magnetization). Y: Yes, N: No, S: Semiconductor, ?: controversial. It is worth noting that $MnTe_2$ can generate SPC and PSC but cannot realize spin-conserved SPC and PSC as in collinear $Rb_{1-\delta}V_2Te_2O$ and $RuO_2$ with AFM magnetic structure.

## II. Symmetry, lattice, magnetism, and band structure

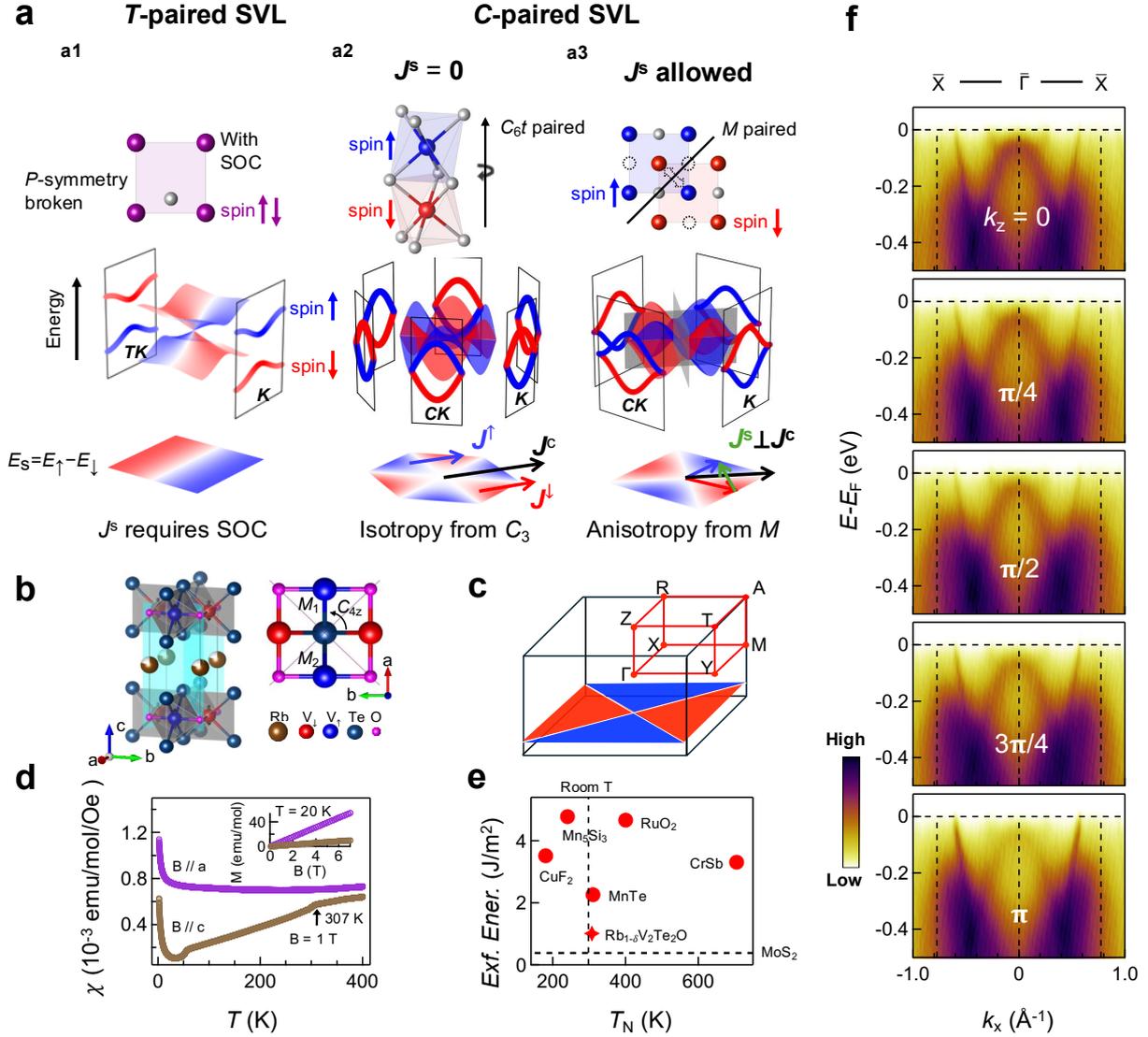

**Fig. 1 *C*-paired SVL in room-temperature AFM Rb$_{1-\delta}$V$_2$Te$_2$O with layered lattice and two-dimensional band structure. a**, Spin splitting and spin (electric) current $J^s$ ($J^c$) in SVL systems: **(a1)**, Spin splitting in *T*-paired SVL systems, where SOC is required, and spin splitting is opposite at *T* connected momentum: $E_\uparrow(K) = E_\downarrow(-K)$. **(a2)**, Type-I *C*-paired SVL systems with isotropic conductance for both spin channels, leading to non-polarized spin current (due to $C_3$ and $C_6t$ symmetry here). **(a3)**, Type-II *C*-paired SVL systems with anisotropic conductance for both spin channels, and pure $J^s$ can be achieved by electric field parallel to the mirror plane. $J^{\uparrow(\downarrow)}$ represent current of spin-up (down) polarized electrons, with $J^s = J^\uparrow - J^\downarrow$ and $J^c = J^\uparrow + J^\downarrow$. **b**, Schematic lattice structure of Rb$_{1-\delta}$V$_2$Te$_2$O, where arrows indicate the magnetic moments of V which are parallel/antiparallel to the *c* axis. Two magnetic sublattices are related by diagonal mirror symmetry $M_1$ and $M_2$ (transparent blue planes) and four-fold rotation symmetry ($C_{4z}$). **c**, Corresponding bulk Brillouin zone (BZ)

with blue and red regions indicating opposite sign of spin polarization among adjacent valleys. **d**, Magnetic susceptibility ($\chi$) vs temperature (T) for magnetic field (1 T) parallel to the *a* and *c* axis. Inset shows the corresponding magnetization (*M*) measured at 20 K. Black arrow indicates one AFM transition around 307 K. **e**, Calculated exfoliation energy (*Exf. Ener.*) and experimentally determined AFM Néel temperature $T_N$ for typical altermagnetic material candidates. The vertical dashed line indicates room temperature, and the horizontal one represents the exfoliation energy of MoS$_2$ as a reference. **f**, ARPES band spectra along the $\overline{X}$-$\overline{\Gamma}$-$\overline{X}$ direction measured at various $k_z$ planes.

Rb$_{1-\delta}$V$_2$Te$_2$O crystallizes in a tetragonal structure with space group *P*4/mmm (No. 123). The single crystal growth method and structure characterization results have been discussed in detail in our previous work[42,43] and are presented here in Methods and Supplementary Section II. The lattice of Rb$_{1-\delta}$V$_2$Te$_2$O consists of layered Rb-Te-V$_2$O-Te slabs in which face-sharing VTe$_4$O$_2$ octahedra are separated by Rb layers (Fig. 1b). The Rb atoms can be deintercalated via a topochemical process[43] and a new compound V$_2$Te$_2$O can be obtained. The van de Waals nature of interlayer coupling suggests that Rb$_{1-\delta}$V$_2$Te$_2$O can be easily exfoliated into ultrathin or even monolayer films. As shown in Fig. 1e and Supplementary Section III, the calculated exfoliation energy, defined as the energy density needed to remove one monolayer from the bulk, is ~1 J/m$^2$ for Rb$_{1-\delta}$V$_2$Te$_2$O, comparable with that of typical layered transition metal dichalcogenide, but much smaller than that of other altermagnetic candidates such as MnTe and RuO$_2$. The layered lattice also results in band structure with negligible $k_z$ dispersion. As shown in Fig. 1f, ARPES spectra along the $\overline{X}$-$\overline{\Gamma}$-$\overline{X}$ direction measured at various $k_z$ planes show no detectable change across the whole bulk BZ, establishing an effectively two-dimensional band structure (more details in Supplementary Section IV).

We have also calculated the ground state energy at various magnetic configurations. As discussed in Supplementary Section V, Rb$_{1-\delta}$V$_2$Te$_2$O favors an AFM configuration in the *ab* plane and ferromagnetic configuration along the *c* axis, with the V magnetic moments parallel/antiparallel to the *c* axis. As shown in Fig. 1b, the two magnetic sublattices are related by mirror symmetries ($M_1$ and $M_2$) and four-fold rotation symmetry ($C_{4z}$) in the Te-O-Te plane but cannot be transformed to each other by any translation operation. Such symmetry lifts spin degeneracy and hence enables the *C*-paired SVL even without SOC[10]. The AFM order in Rb$_{1-\delta}$V$_2$Te$_2$O is verified experimentally by the magnetic susceptibility ($\chi$) characterization. As shown in Fig. 1d, both curves for $B \parallel a$ and $B \parallel c$ show very weak susceptibility in the order of

$10^{-3}$ emu /(mol·Oe), as expected from the compensated moments of AFM order. The low-temperature (<20 K) upward tails of the $\chi$-$T$ curves are attributed to the paramagnetic impurities[42]. At the high temperature region (100-400 K), while the $\chi$-$T$ curve for $B \parallel a$ exhibits temperature independent behavior, the $\chi$-$T$ curve for $B \parallel c$ keeps a negative slope in the whole range and further develops a slope jump at ~ 307 K, signifying an overall AFM phase and phase transition at $T_N$ ~ 307 K. The magnetic field dependent magnetization curve (inset of Fig. 1d) shows no trace of moment rearrangement up to $\mu_0 H = 7$ T. Due to the instrumental limitation, the existence of phase transition from paramagnetic phase to AFM phase remains to be clarified for the temperature range above 400 K. These suggests that $Rb_{1-\delta}V_2Te_2O$ belongs to one of the room-temperature altermagnets such as $\alpha$-$MnTe$[22,23] and $CrSb$[11,24] (Fig. 1e).

Due to the ferromagnetic configuration along the $c$ axis, there is no spin reversal between $\pm k_z$ (Supplementary Section VI). Together with the layered lattice, the band structure and spin polarization of $Rb_{1-\delta}V_2Te_2O$ can be discussed simply in the arbitrarily projected BZ. Figure 1c depicts such projection with bule and red colors denoting the alternating spin splitting locked to momentum due to the crystal symmetry. In general, our lattice and band structure characterizations establish $Rb_{1-\delta}V_2Te_2O$ as a layered, room-temperature AFM materials with $C$-paired SVL, rendering it outstanding from other altermagnetic candidates, facilitating the potential application in AFM spintronics, valleytronics and other novel applications associated with two-dimensional materials.

## III. Measured/calculated band structure and temperature dependence

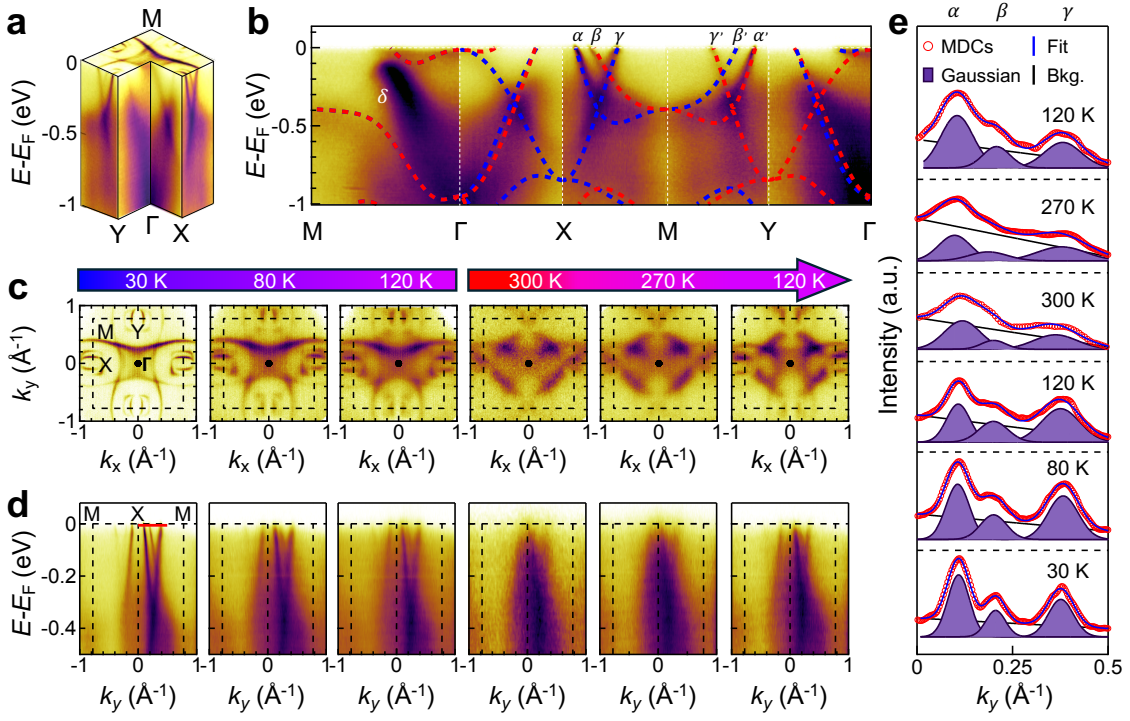

**Fig. 2 Giant spin splitting persisting up to room temperature. a**, ARPES spectral intensity plotted in $E_B$-$k_x$-$k_y$ space. **b**, Band spectra along high-symmetry paths overlaid with results from first-principles calculations. Red and bule dashed lines indicate bands with opposite sign of spin. Bands denoted as $\alpha$, $\beta$, $\gamma$ and $\alpha'$, $\beta'$, $\gamma'$ are related with diagonal mirror symmetry and spin splitted by $C$-paired AFM order. Band $\delta$ is attributed to surface reconstruction of Rb atoms, see Supplementary Section VII for details. **c-d**, Temperature dependent Fermi surfaces (**c**) and spectra along M-X-M path (**d**). **e**, Corresponding integrated momentum distribution curves (MDCs) at the Fermi level (±20 meV). The three peaks correspond to $\alpha$, $\beta$, and $\gamma$, respectively, as shown at the top of the panel. The MDCs representing the original spectral density (red circles) are fitted by blue lines composed of multiple Gaussian peaks (filled blue peaks) and a linear background (black lines). Here all the ARPES results are measured using a photon energy $h\nu = 69$ eV, corresponding to $k_z = 6 \times 2\pi/c$.

Figure 2a and b present the ARPES spectra in $E_B$-$k_x$-$k_y$ space and high-symmetry spectra measured at 12 K, overlaid with calculated bands. Three key features can be clearly extracted, which directly reflect the lattice and magnetic symmetry. Firstly, although with intensity inhomogeneity, the Fermi surface hold four-fold rotational symmetry in $k_x$-$k_y$ plane, in line with the tetragonal lattice; The second in-plane symmetric operation is the diagonal mirror symmetry ($M_1/M_2$) which connects the two magnetic sublattices in real space and the adjacent valleys in reciprocal space. These operations result in symmetric band spectra along X-Γ-Y and X-M-Y (Fig. 2b), respectively; Lastly, the low-energy band structure and Fermi surface are dominated by three sharp bands $\alpha$, $\beta$ and $\gamma$. The colored dashed lines from first-principles calculations for the AFM state exhibit not only the same sign of spin polarization between bands $\alpha$ and $\gamma$ while the opposite for band $\beta$, but also the alternating spin polarization between X-M and Y-M. The great match between first-principles calculations and ARPES measurement fully establishes the C-paired SVL in $Rb_{1-\delta}V_2Te_2O$.

We have also performed temperature dependent measurement for both Fermi surfaces (Fig. 2c) and band spectra (Fig. 2d) with temperature increasing from 30 K to 300 K, and then cooling down again. At low temperature (30 K), bands $\alpha$, $\beta$ and $\gamma$ are clearly resolved, manifested as closed pockets surrounding X/Y valleys in the Fermi surface mapping and sharp peaks in the MDCs (Fig. 2e). With increasing temperature, these band features become blurred with weakened peak intensity and broadened peak width. Approaching 300 K, the MDC peaks corresponding to bands $\alpha$ and $\gamma$ remain robust, the middle peak from band $\beta$ becomes ill-defined in the raw MDC but can still be resolved by fitting with multiple Gaussian peaks, suggesting the robustness of spin splitting. These peaks regain their sharpness when cooling back to low temperature. Altogether, such results demonstrate the persistence of SVL up to room temperature.

## IV.    Spin resolved ARPES spectra

The alternating spin polarization of C-paired SVL can be directly demonstrated via image-type Spin-ARPES[41] measurement on the photoelectron spin polarization. As schematically illustrated in Fig. 3a, the Fe target (spin filter) can be magnetized by a Helmholtz coil along the $\pm x$ or $\pm y$ direction. Correspondingly, the spin components $S_x$ or $S_y$ of the photoelectrons can be detected. To reach the X/Y valley around the BZ boundary, the single crystal was rotated along the $y$ axis by a certain angle $\theta$. In this case, while the measured $S_y$ comes from the initial $S_y$ polarization of electrons inside the crystal, the measured effective $S_x$ corresponds to the projection of initial

$S_x$ and $S_z$ to the $x$ axis, i.e., $S_{x\text{ eff.}} = S_x\cos\theta + S_z\sin\theta$.

Figure 3b and 3c compare the calculated spin resolved Fermi surface to the ARPES measured one. In line with the high symmetry path spectra shown in Fig. 2b, the Fermi surface consists of three pockets $\alpha$, $\beta$ and $\gamma$ surrounding X/Y. In the current AFM configuration, the spin of these low-energy bands is polarized along the $z$ direction. As shown in Fig. 3b, the inner and outer pockets $\alpha$ and $\gamma$ share the same sign of spin polarization $+S_z$, while the middle pocket $\beta$ has opposite direction $-S_z$. Due to the C-paired SVL, the spin polarization of these bands has opposite sign for adjacent valley, i.e., $-S_z$ for $\alpha'$, $\gamma'$ and $+S_z$ for $\beta'$ surrounding the Y valley.

We select the most typical spectral cuts related by the mirror symmetry, cut 1 along $\overline{\text{X}}$-$\overline{\text{M}}$ and cut 2 along $\overline{\text{Y}}$-$\overline{\text{M}}$, to directly prove the C-paired SVL. Figure 3d and 3e present the Spin-ARPES results containing spin resolved spectral image, spin-resolved MDCs, and spin polarization component for $S_{x\text{ eff.}}$. The data sets corresponding to spin polarization component $S_y$ are shown in Supplementary Section VI. First of all, we rule out the existence of in-plane ($S_x$ and $S_y$) spin component from the measured results. As shown in Fig. S10 and S13, the spectral cuts show no detectable polarization of $S_y$, as judged by the absence of polarization peaks corresponding to the band features. Note the sharp peaks in these curves (Fig. S13) come from the system noise due to the notoriously low efficiency of Spin-ARPES but does not correspond to any real band features. This suggests that neither $S_y$ for X valley nor $S_x$ for Y valley (tangential spin texture) could exist, since $S_x$ will transfer to $S_y$ in the process of rotating the sample azimuth from measuring cut 1 to cut 2. Furthermore, in Fig. S14, Spin-ARPES results using a single-channel VLEED spin detector are presented. Such detector and geometry could measure the spin components $S_z$ and $S_y$ directly. Here, only $S_z$ polarization can be resolved while no sign of $S_y$ corresponding to the band features at Y valley (radial spin texture) was detected. Consequently, Spin-ARPES measurements using two different geometries provide complementary and consistent results, demonstrating the absence of in-plane ($S_x$ and $S_y$) spin polarization. We thus conclude that in Fig. 3 the measured $S_{x\text{ eff.}}$ reflects the projection of $S_z$ on the $x$ axis ($S_z = S_{x\text{ eff.}}/\sin\theta$), in agreement with the first-principles calculation shown in Fig. 3b.

Having established the existence of $S_z$ polarization, we then examine its momentum distribution. As shown in Fig. 3d1 by spin-resolved spectra cut 1, bands $\alpha$ and $\gamma$ have $+S_z$ while $\beta$ has $-S_z$. Spin resolved MDCs and polarization curves shown in Fig. 3d2 also present broad peaks (indicated by blue and red rectangles), corresponding to the real band features in the right energy and momentum region. Moving to cut 2, the $S_z$ polarization remains detectable (although weaker due to surface aging), and moreover, the sign of polarization is indeed reversed for all the three bands. Now, bands $\alpha'$ and $\gamma'$ have $-S_z$ while $\beta'$ has $+S_z$ polarization. These results well fit the theoretical prediction shown in Fig. 3b, providing defining evidence for the alternating spin polarization in adjacent valleys as a manifestation of C-paired SVL.

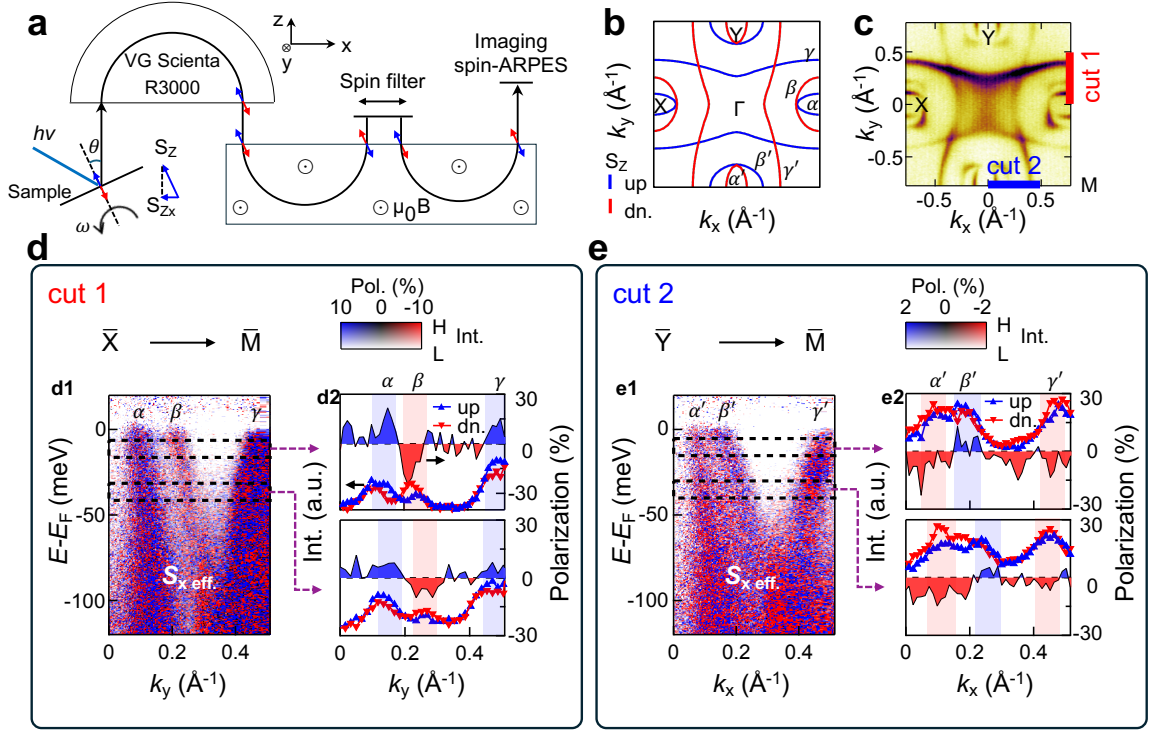

**Fig. 3 Observation of C-paired SVL with alternating sign of spin polarization $S_z$. a**, Schematic top view of image-type spin-ARPES system and the measurement geometry. **b**, Spin-resolved Fermi surface for the AFM phase from the first-principles calculations. **c**, ARPES Fermi surface mapping ($h\nu = 69$ eV). Red and blue solid lines (cut 1 and cut 2) indicate the momentum path of ARPES spectra in **d** and **e**, respectively. **d**, Spin-resolved band spectra of cut 1 along the path $\overline{X}$-$\overline{M}$. **d1**, Spin-imaged spectra of polarization $S_{x\,\text{eff.}}$. **d2**, Spin-resolved MDCs (left axis) and corresponding spin polarization $S_{x\,\text{eff.}}$ (right axis) integrated from energy windows indicated by the horizontal dashed boxes in **d1**. **e**, same as **d** but for cut 2 along the path $\overline{Y}$-$\overline{M}$. Spin-ARPES spectra in **d** and **e** are taken with photon energy $h\nu = 21.2$ eV. Transparent red and blue rectangles indicate the spin polarization peaks coming from the real band features.

# V. QPI pattern

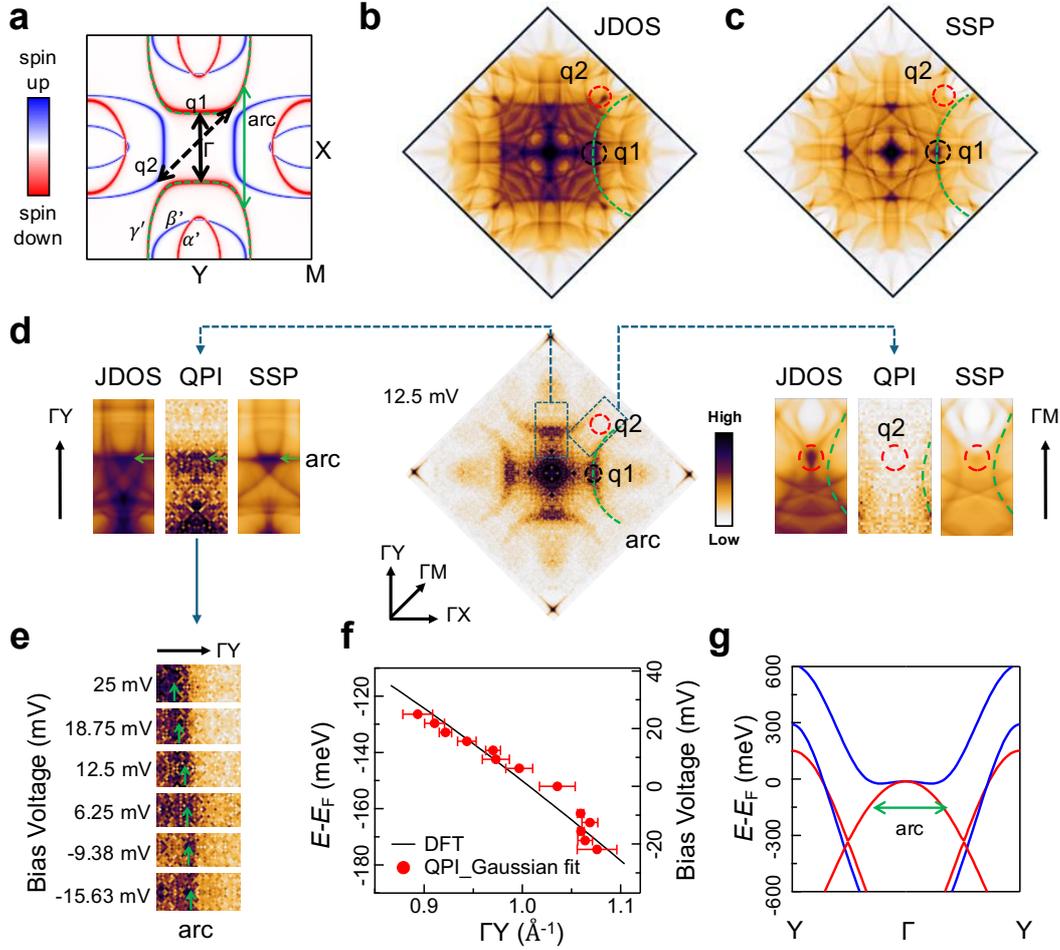

**Fig. 4 Suppression of intervalley scattering observed in the QPI pattern. a**, Calculated constant energy contour at -100 meV. Black double arrows mark the scattering wavevectors q1 and q2. Green double arrow indicates the $\gamma' - \gamma'$ scattering which forms the arc shown in **d**. **b,** Calculated JDOS at $E$-$E_F$ = -140 meV. **c,** Same as **b** but with the consideration of spin-dependent scattering probability, SSP. **d,** Middle panel shows the FFT of STS $dI/dV$ map at $V_{bias}$ = 12.5 mV, $I$ = 50 pA, for a 45 × 45 nm region. Green dashed curve indicates the convex arc, while q1 (q2) is indicated by the black (red) dashed circle. The left and right panels show the zoom-in of QPI patterns along ΓY and ΓM, respectively, in comparison with the JDOS and SSP simulation. **e,** Cut slices of QPI along ΓY at different bias voltage ranging from -15.63 mV to 25 mV. **f**, Energy dispersion of q1 from QPI and its comparison with first-principles calculations. The positions of q1 in QPI are determined through Gaussian fittings. In both **e** and **f** the Fermi level is aligned to the ARPES result. **g,** Calculated band structure along the Y-Γ-Y direction.

In addition to ARPES measurements, STM/STS extends the investigation to both occupied and unoccupied states in real and reciprocal space, aiding in the study of alternating SVL. Elastic scattering from disorder/defects can mix the electronic eigenstates across reciprocal space, resulting in QPI patterns at wavevector $\boldsymbol{q} = \boldsymbol{k_1} - \boldsymbol{k_2}$ with $\boldsymbol{k_{1/2}}$ the moments. Given that scattering between bands with misaligned spin components can be suppressed by the spin selection rule, QPI patterns can reveal spin characteristics within the electronic structure[44,45].

Firstly, we calculate the momentum-resolved density of states $\rho$, as shown in Fig. 4a, where red (blue) represents spin down (up) polarized bands. To elucidate the $k$ space contributions from these bands, we simulate the joint density of states (JDOS) by $\mathrm{JDOS}(\boldsymbol{q}, E) = \int_{\mathrm{BZ}} \rho(\boldsymbol{k}, E)\rho(\boldsymbol{k} + \boldsymbol{q}, E)\mathrm{d}^2 k$, without considering spin-dependent scattering processes. As shown in Fig. 4a and Fig. 4b, we focus on three types of scattering patterns: wavevector q1 along the ΓY direction, resulting from scattering of valleys with the same spin; wavevector q2 along the diagonal direction, resulting from the scattering of valleys with opposite spin; and the arc representing inter-$\gamma'$ band scattering, plotted as dashed green lines in Fig. 4b. The detailed analysis of all the eight JDOS scattering patterns are presented in Supplementary Section VIII. We further simulate QPI pattern considering spin-dependent scattering probability (SSP). As shown in Fig. 4c, SSP patterns retain the q1 and arc features, while attenuate near q2, because of the spin selection rule.

To experimentally investigate the QPI patterns, we performed STS measurements at the Te-terminated surface. Upon removal of Rb atoms on the freshly cleaved surface by STM tip, we reach the Te termination with tetragonal $1 \times 1$ lattice (see Supplementary Section VII). The QPI patterns shown in the middle panel of Fig. 4d are the fast Fourier transform (FFT) of a d$I$/d$V$ map at Te termination. Such FFT QPI patterns reveal four prominent arcs (green dashed curves) and strong intensity peaks near the center of these arcs (black dashed circles as q1). Both the arc and q1 features are consistent with the JDOS (Fig. 4b) and SSP (Fig. 4c) simulations. Conversely, q2 feature is absent in the experimental QPI pattern (right panel of Fig. 4d), in line with the SSP simulation. The suppression of q2 indicates the prohibition of scattering between opposite spin polarized states, further demonstrating opposite spin polarization in crystal-symmetry connected valleys.

Figure 4e shows the QPI patterns as a function of energy, displaying a progressive inward shift of these arcs with increasing bias voltage. This suggests a negative slope of the corresponding wavevector, consistent with bands from first-principles calculations (Fig. 4g). Figure 4f presents the energy dispersion of the QPI arc, which aligns well with the JDOS calculation when a downward ~150 meV energy shift is applied to the latter, anticipated due to the removal of Rb atoms during the STM measurement. Furthermore, by normalizing the intensity of q1 on the arc in the JDOS, SSP and QPI patterns, we observe an overall decreased signal in QPI and SSP compared to JDOS. Such reduction is also consistent with the intrinsic spin polarization of the bands, as spin selection rules dominate the scattering process on the surface.

In summary, the agreement between the SSP and QPI pattern, particularly the attenuation of the q2 wavevector, provides direct physical consequence of *C*-paired SVL.

## VI. Discussion

In conclusion, combining ARPES, spin-ARPES, STM/STS measurements, and first-principles calculations, we provide direct evidence for the realization of *C*-paired SVL in a layered, room-temperature AFM $Rb_{1-\delta}V_2Te_2O$ single crystal. We not only prove the existence of alternating sign of spin splitting in adjacent valleys of the tetragonal BZ using Spin-ARPES measurement, characteristic of *C*-paired SVL, but also demonstrate the direct physical response of such SVL, the spin scattering attenuation between adjacent valleys, via QPI analysis. The realization of such *C*-paired SVL in real, layered $Rb_{1-\delta}V_2Te_2O$ single crystals provide a unique material platform that allows noncollinear nonrelativistic spin current and unconventional piezomagnetism, holding significant potential in developing new physics and effect in AFM materials. In addition, with SOC and/or the proximity effect further considered, many other exotic phases like topological or finite-momentum superconducting states can be expected for fundamental research, and various transport properties like anomalous Hall[46] and Nernst effect[47] or nonlinear Hall effect can be realized and used to detect or manipulate the Néel vector[48].

## VII. Methods

**Sample growth and characterization**

Single crystals of $Rb_{1-\delta}V_2Te_2O$ were grown via spontaneous nucleation using the self-flux RbTe[42,43]. The source materials include Rb (99.5%), $V_2O_5$ (99.99%), Te (99.999%), and V (99.5%), and the overall composition was Rb:V:Te:O = 6:2:7:1 in molar ratio. The raw materials were mixed, loaded into an alumina crucible, and then sealed in an evacuated silica ampule. The ampule underwent heating at 300 °C for 1000 minutes to facilitate pre-reactions involving the reactive Rb metal. After the first stage of the pre-reactions, the sample was taken out and put into an argon-filled glove box (with oxygen and water content below 1 ppm), and quickly homogenized by grinding. Subsequently, the ground powders were reloaded into the alumina crucible, sealed in a Ta tube to prevent reactions with quartz, and placed inside an evacuated silica ampule. The assembly was gradually heated to 1000 °C for over 1000 minutes in a muffle furnace, followed by cooling down to 950 °C within 100 minutes. The temperature was set to decrease at a rate of 2 °C /h from 950 °C to 650 °C. At the end of the crystal growth, the sample assembly underwent centrifugation to separate the grown crystals from the flux.

The structure of the crystals was determined by x-ray diffraction using Cu $K_\alpha$ radiation at room temperature with a PANalytical diffractometer (Empyrean Series 2), and Mo $K_\alpha$ radiation with a Bruker D8 VENTURE at 100 K. The diffraction patterns can be well indexed by the (*00l*) and (*hk0*) reflections. Magnetization measurements were performed using Magnetic Property Measurement System (MPMS3, Quantum Design). X-ray absorption near edge structure spectra were performed at BL07U of the Shanghai Synchrotron Radiation Facility.

**ARPES and spin-resolved ARPES measurements**

ARPES measurements were performed at the BL03U beamline of the Shanghai Synchrotron Radiation Facility equipped with a Scienta Omicron DA30 energy analyzer and *p*-polarized radiation. The samples were cleaved in situ under base pressure better than $5\times10^{-11}$ mbar and temperature below 15 K and were cleaved at 300 K during data collection for the cooling measurement. The $k_z$ dispersion was measured using photon energies between 50 and 100 eV. During the experiment, the beam spot size was set to 15×15 μm². The energy resolution was set to 10-20 meV depending on the photon energy used, and the angular resolution was set to 0.2°.

The spin-ARPES measurements were performed using a VG Scienta R3000 analyzer integrated with a homemade multichannel very-low-energy-diffraction (MCVLEED) spin polarimeter[41] and a He I$\alpha$ (21.2 eV) light source. The energy and angular resolutions are 12 meV and 0.5°, respectively. The samples were cleaved *in situ* and measured at $5\times10^{-10}$ mbar and 6 K. Spin-APRES measurements were also performed in the BL9B of Hiroshima Synchrotron Radiation Center, Japan and BL13U of National Synchrotron Radiation Laboratory, Hefei.

**First-principles calculations**

The calculations were performed in the framework of density functional theory as implemented in Vienna ab initio simulation package[49]. The projector-augmented wave potential was adopted with the plane-wave energy cutoff set at 440 eV. The exchange-correlation functional of the Perdew–Burke–Ernzerhof type has been used for both structural relaxations and self-consistent electronic calculations, with the convergence criteria as $10^{-5}$ eV[49,50]. The GGA+$U$ method was employed to treat the strong correlations of the V 3$d$ orbitals, where the value of the Hubbard $U$ was taken as 1 eV, which provided the best fitting to the ARPES results. The Brillouin zone was sampled by a $7\times7\times5$ Γ-centered Monkhorst–Pack mesh, $41\times41\times1$ k-mesh was used for Fermi surface calculation in the $k_z = 0$ plane. In exfoliation energy calculations, denser k mesh was tested carefully for each material to ensure the convergence. We construct Wannier tight-binding model Hamilton by the WANNIER90 interface[51], with V 3$d$ orbitals and Te 5$p$ orbitals, the density of states $\rho$ in JDOS and SSP was calculated with obtained Wannier model in $201\times201\times1$ k-mesh.

**Scanning tunnelling microscopy measurements**

STM measurements were conducted using a Unisoku USM-1500 scanning probe microscope system. These measurements were conducted at 4.4 K within an ultra-high vacuum chamber, where the pressure was maintained at approximately $1\times10^{-10}$ mbar. The samples were cleaved in situ at 77 K within the vacuum chamber and immediately transferred to the sample stage without exposure. d$I$/d$V$ spectroscopy data were obtained using a Nanonis controller equipped with a built-in lock-in amplifier. The lock-in amplifier was set to operate at a modulation frequency of 971.9 Hz and an amplitude of 5 mV. JDOS patterns were generated by self-convolution of the calculated constant energy contours (CECs). These CECs were aligned with experimental results from ARPES. SSP patterns were derived by self-convolving bands with the same spin polarization separately and subsequently summing these results, as the fully spin polarized bands precludes scatterings between antiparallel spins.

**Data availability**

All data are available in the main text or the supplementary materials. Further data are available from the corresponding author on reasonable request.

**ACKNOWLEDGEMENTS**

We thank Dr. Ruixin Guo, Dr. Shu Guo, Dr. Tengyu Guo, Dr. Qi Liu for the assistance in sample characterization, Dr. Liusuo Wu, Dr. Jieming Sheng, Dr. Enke Liu, Dr. Xiaoming Ma, Dr. Bing Shen. Dr. Yuan Wang, Dr. Wanling Liu, Dr. Wencheng Huang, Dr. Zhanyang Hao for helpful discussion, Dr. Jianyang Ding, Dr. Weihong Sun, Dr. Yamei Wang, Dr. Fangyuan Zhu, Dr. Yuliang Li, Dr. Yi Liu, Dr. Koji Miyamoto and Dr. Kazuki Sumida for beamline support.

This work is supported by the National Key R&D Program of China (Grants No. 2021YFA1401500, 2022YFA1403700, and 2022YFA1403202), the National Natural Science Foundation of China (NSFC) (Grant No. 12074163), Hong Kong Research Grants Council (Grants No. 16303821, 16306722, and 16304523), Guangdong Basic and Applied Basic Research Foundation (Grants No. 2022B1515020046, 2022B1515130005 and 2021B1515130007), the Guangdong Innovative and Entrepreneurial Research Team Program (Grant No. 2019ZT08C044), Shenzhen Science and Technology Program (Grant No. KQTD20190929173815000).

**Author contributions**

C.C. and J.L. conceived the idea and proposed the experimental and modelling design. F.Z., L.D., Y.Q., Z.L., M.Y., Y.H., C.Z., T.O., K.S., S.C. and S.Q. contributed to the (spin-)ARPES instruments, measurement and analysis. F.Z., C.L., J.L., X.M. and G.C. contributed to the sample growth and characterization. X.C., Z.S. and J.L. performed calculations/simulations. Z.Y., S.Z. and Y.Z. performed the STM/STS measurements and analysis. All authors wrote and corrected the manuscript.

**Competing interests**

The authors declare no competing interests.

**Correspondence and requests for materials** should be addressed to Junwei Liu or Chaoyu Chen